\documentclass[10pt,twocolumn]{article}
\usepackage{times,amsmath,epsfig}

\makeatletter
\newcommand{\tinyspacing}{\let\CS=\@currsize\renewcommand
{\baselinestretch}{0.97}\tiny\CS}
\newcommand{\singlespacing}{\let\CS=\@currsize\renewcommand
{\baselinestretch}{1}\tiny\CS}
\newcommand{\myspacing}{\let\CS=\@currsize\renewcommand
{\baselinestretch}{1.2}\tiny\CS}
\makeatother

\tinyspacing

\usepackage{amsmath,amssymb}
\usepackage[boxed]{algorithm}
\usepackage{algorithmic}
\usepackage{graphicx}
\usepackage{multirow}
\usepackage[tight]{subfigure}
\usepackage[square,sort,comma,numbers]{natbib}

\usepackage{balance}
\usepackage{url}
\usepackage{multirow}
\usepackage{latexsym}
\usepackage{booktabs}

\usepackage{fancyvrb}
\usepackage{dsfont}
\usepackage{color}

\renewcommand{\paragraph}[1]{\vspace{1mm}\noindent{\bf #1}}
\newcommand{\eat}[1]{}

\newcommand{\cobbler}{{CBLOCK}}
\newcommand{\btree}{{BlkTree}}

\newcommand{\rbox}{\hfill $\Box$}

\newtheorem{definition}{Definition}[section]

\newtheorem{lemma}[definition]{Lemma}

\newtheorem{theorem}[definition]{Theorem}

\newtheorem{example}[definition]{Example}

\newcommand{\squishlist}{
   \begin{list}{$\bullet$}
    { \setlength{\itemsep}{-1pt}      \setlength{\parsep}{3pt}
      \setlength{\topsep}{1pt}       \setlength{\partopsep}{0pt}
      \setlength{\leftmargin}{1.5em} \setlength{\labelwidth}{1em}
      \setlength{\labelsep}{0.5em} } }

\newcommand{\squishlisttwo}{
   \begin{list}{$\bullet$}
    { \setlength{\itemsep}{0pt}    \setlength{\parsep}{0pt}
      \setlength{\topsep}{0pt}     \setlength{\partopsep}{0pt}
      \setlength{\leftmargin}{2em} \setlength{\labelwidth}{1.5em}
      \setlength{\labelsep}{0.5em} } }

\newcommand{\squishend}{
    \end{list}  }

\title{CBLOCK:  An Automatic Blocking Mechanism for Large-Scale De-duplication Tasks}
\author{
Anish Das Sarma{\small $~^{\#1}$}, Ankur Jain{\small $~^{*2}$}, Ashwin Machanavajjhala{\small $~^{*2}$}, Philip Bohannon{\small $~^{*2}$}\\
{\small $~^{*1}$}Google Research, {\small $~^{*2}$}Yahoo! Research\\
\{anish.dassarma\}@gmail.com, \{ankurja,mvnak,plb\}@yahoo-inc.com\\
}
\date{}

\begin{document}
\maketitle

\begin{abstract}

De-duplication---identification of distinct records referring to the same real-world entity---is a well-known challenge in data integration. Since very large datasets prohibit the comparison of every pair of records, {\em blocking} has been identified as a technique of dividing the dataset for pairwise comparisons, thereby trading off {\em recall} of identified duplicates for {\em efficiency}. Traditional de-duplication tasks, while challenging, typically involved a fixed schema such as Census data or medical records.  However, with the presence of large, diverse sets of structured data on the web and the need to organize it effectively on content portals, de-duplication systems need to scale in a new dimension to handle a large number of schemas, tasks and data sets, while handling ever larger problem sizes.  In addition, when working in a map-reduce framework it is important that canopy formation be implemented as a {\em hash function}, making the canopy design problem more challenging. We present \cobbler, a system that addresses these challenges.

\cobbler\ learns hash functions automatically from attribute domains and a labeled dataset consisting of duplicates. Subsequently, \cobbler\ expresses blocking functions using a hierarchical tree structure composed of atomic hash functions. The application may guide the automated blocking process based on architectural constraints, such as by specifying a maximum size of each  block (based on memory requirements), impose disjointness of blocks (in a grid environment), or specify a particular objective function trading off recall for efficiency. As a post-processing step to automatically generated blocks, \cobbler\ {\em rolls-up} smaller blocks to increase recall. \eat{Finally, instead of using a labeled dataset, \cobbler\ can also be used to automatically block a dataset based on a given set of {\em match-rules} used for de-duplication. }We present experimental results on two large-scale de-duplication datasets at Yahoo!---consisting of over 140K movies and 40K restaurants respectively---and demonstrate the utility of \cobbler.



\end{abstract}

\section{Introduction}
\label{sec:intro}
\label{sec:introduction}


Integrating data from multiple sources containing overlapping information invariably leads to {\em duplicates} in the data, arising due to different sources representing the same entities (or facts) in slightly different ways; e.g., one source says ``George Timothy Clooney'' and another says ``G. Clooney''. The problem of identifying different records referring to the same real-world entities is known as {\em de-duplication}\footnote{De-duplication is also known by many other names such as reference reconciliation, record linkage, and entity resolution.}. De-duplication has been identified as an important problem in data integration, and has enjoyed significant research interest, e.g.~\cite{er59,fsunter69,sb02,tkm01,eiv07,gbvr03,winkler06}.

Conceptually, de-duplication may be performed by considering each pair  of records, and applying some {\em matching function}~\cite{koudas06, secondstring,fsunter69} to compute a similarity score, then determining duplicate sets of records based on clustering similar pairs. However, comparing all pairs of records to be de-duplicated is prohibitively expensive in commercial or web applications that require matching data sets with millions of records (e.g., persons, business listings, etc). {\em  Blocking} or {\em canopy-formation} (e.g.,~\cite{baxter,merge-purge,kelley85,jaro89,swoosh-blocking,canopy-kdd,bilenko06:blocking,knoblock}) has been identified as a standard technique for scaling de-duplication: The basic idea is to find a set of (possibly overlapping) subsets of the entire dataset (called {\em blocks}), and then compute similarity scores only for pairs of entities appearing in the same block. We use the term ``blocking function'' to refer to any function that maps entities to block numbers, usually based on the value of one or more attributes.  One example of a blocking function would be the value of the ``phone number'' attribute, or the first seven digits of the same, etc.
 In an ideal situation, all (or most) of the duplicates would appear together in at least one block.

As a result, a good blocking function must be designed for {\em each large-scale matching task}.
\eat{,  and recent work~\cite{bilenko06:blocking,knoblock} has investigated the problem of {\em automatically designing} blocking functions for new de-duplication tasks.  The starting point of this work is a set of  {\em positive and negative examples} of matching pairs for~\cite{bilenko06:blocking} and {\em positive examples} for~\cite{knoblock}.  These training samples may already exist because the data set contains some strong keys (e.g. ``ISBN number''), or may be editorially created.  Using this training set and a library of hash functions, a set of functions or conjunctions of functions is found that maximizes the number of positive matches covered while minimizing the number of negative matches, without using ``too many'' blocking functions~\cite{bilenko06:blocking}. The solution of~\cite{knoblock} is similar, except it only considers {\em positive examples}. }
We are seeking to build a scalable system for de-duplication of web data. The system will be used for a wide variety of de-duplication tasks, and must support {\em agility}, the ability to rapidly develop new de-duplication applications.  Accordingly, an important part of developing this system is effective, automatic construction of blocking functions.
Like~\cite{vernica10}, de-duplication tasks in our system execute in a map-reduce framework like Hadoop.  In this setting, computation is broken into rounds consisting of a {\em map phase} in which a set of keys is generated by which work is split over a potentially large number of compute nodes and a {\em reduce phase} in which partial results from each compute node are combined.  A natural approach for de-duplication is to use the map phase to execute the blocking function, allowing match scores to be computed in parallel on each mapper.

In order to design appropriate blocking functions for our setting, we face four important challenges.  First, a premium is placed on minimizing the number of rounds of computation in a map-reduce setting, since each round involves significant scheduling and co-ordination overheads.  Second, data in our system comes from a variety of feeds, and is often noisy.  In particular, attributes may be only partially populated, leading to asymmetric block sizes if these attributes are used for blocking. Third, matching is executed in parallel, meaning that a premium is placed on minimizing the size of the largest block without exceeding the maximum number of compute nodes available.  Fourth, the complexity of the de-duplication process can be significantly reduced if every object is given only a single hash value for mapping; which we refer to as the {\em disjoint} blocking condition.

We present \cobbler, a system that automatically creates canopies based on the information specified by the application.  We now describe the approach taken in \cobbler\ to address the above challenges.
We introduce a conditional tree of blocking functions, the {\em \btree}.  In this tree, blocks with large expected size are explicitly mapped to a child blocking function, making each path in the tree equivalent to a conjunctive blocking function applied to a subset of the data.  The introduction of the \btree\ allows for an expressive blocking function, which allows us to effectively block even skewed data, such as attributes with many null values.

Second, to handle the situation in which the number of blocks exceeds the number of compute nodes, we introduce a {\em roll-up} step for the \btree\ to efficiently reduce the number of compute nodes without excessively increasing complexity of the hash function. Third, we optimize for the best blocking function while keeping the size of the largest block within a constrained size.  Since the overall latency of the parallel computation corresponds to the slowest node, this is a natural optimization goal, but is not addressed by existing techniques.  As an aside, we note that our system can also be used for other applications that require a similar capability as blocking: (a) In a binary classifier with many features, \cobbler\ may be used to pick a small set of features that most effectively captures the classification; (b) We can use \cobbler\ to determine which sets of values from two relations may contribute most to join results in a distributed join solution such as~\cite{thetajoin}.

\begin{figure}[t]
\centering
\includegraphics[width=2.5in]{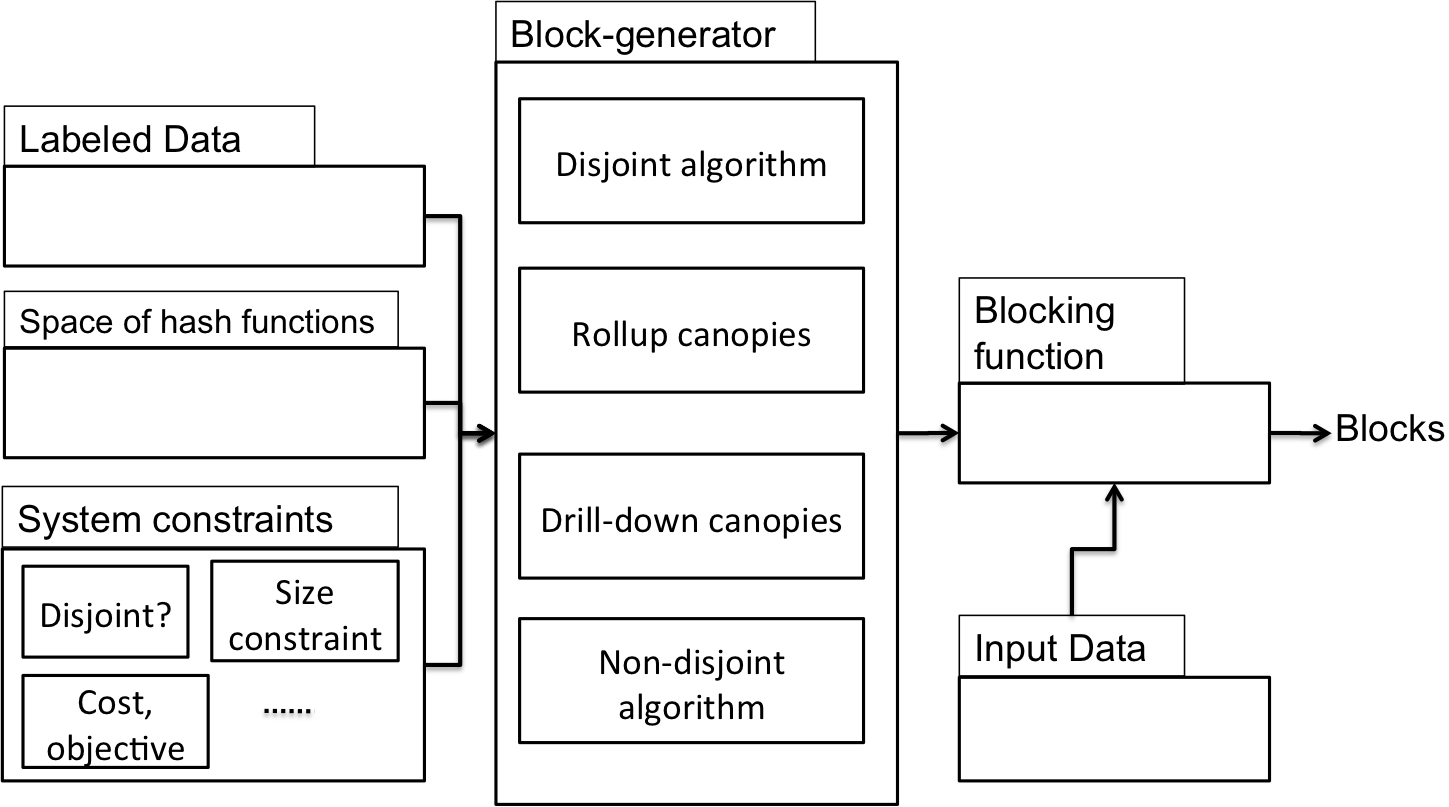}
\vspace{-5mm}
\caption{\label{fig:cobbler}Components of the \cobbler\ system}
\end{figure}

The flow of data through \cobbler\ is illustrated in Figure~\ref{fig:cobbler}. As input to the system, shown on the left, is a set of training examples consisting of true-positive match pairs shown at the top, and a set of configuration parameters shown a the bottom including size constraints, disjointness conditions and any tuning of the cost objective.  These inputs feed into the \cobbler\ system, shown in the middle block, that designs a blocking configuration. This configuration is then passed to the runtime system (e.g. a map-reduce system) for execution of the blocking as the first phase of the de-duplication algorithm.

\subsection{Contributions}
\label{subsec:outline}

Our paper makes the following contributions, addressing the requirements of automatic blocking configuration for web-scale de-duplication:

\squishlist

\item In order to decrease the number of rounds (disjuncts), we argue that it is necessary to increase the power of individual hash functions while still respecting disjointness constraints.
In Section~\ref{sec:labeled} we formally introduce {\em blocking trees} to accomplish this goal.  We show that in general finding blocking trees that maximize recall subject to a {\em maximum block-size} constraint is NP-hard, and we provide a natural greedy algorithm.

\item We show that adapting the state of the art solution of~\cite{bilenko06:blocking,knoblock} to optimize for the {\em maximum size} constraint can naturally be expressed as a special case of finding optimal blocking trees.

\item Section~\ref{sec:rollup} introduces the roll-up problem of merging small canopies produced using any disjoint blocking scheme. We show a close connection of the roll-up problem with the knapsack problem, establish the NP-completeness of solving the general problem, and provide a heuristic algorithm based on a 2-approximate algorithm for the knapsack problem.

\item Section~\ref{sec:hashsearch} studies ``drill-down'' problem, i.e., given a domain of an attribute and a labeled dataset of true duplicates we want to find optimal hash functions that meet a canopy-size requirement. We formally define the problem and present a near-linear time optimal algorithm based on dynamic programming.

\item For most of the paper we focus our attention on disjoint blocking functions. Section~\ref{sec:nondisjoint} extends our study to non-disjoint blocking functions. To our knowledge, this is the first work to consider the disjointness issue for blocking design.

\eat{
}

\item \cobbler\ is fully implemented along with all the functionality described above. In Section~\ref{sec:experiments}, we describe our system and present  experimental results on two large commercial datasets consisting of around 140K movies and 40K restaurants respectively.

\eat{
\item {\bf [[Remove this assuming we are dropping these points.]]} In Section~\ref{sec:futurework} we present some initial ideas on other topics of blocking.In Section~\ref{sec:rules} we turn to the discussion of match-rules as a method of providing information on recall of blocking techniques. We define a formal language for specifying matching rules, define the optimization problem of finding blocking based on match rules, and provide optimal algorithms. In Section~\ref{sec:mechanisms} we introduce a space of (not necessarily disjoint) canopy formation mechanisms, and express trade-offs between them based on various desirable properties.
}
\squishend

\noindent Related work is described in Section~\ref{sec:relatedwork}.
Due to space constraints, formal proofs for all technical results are omitted from the paper.

\section{Related Work}
\label{sec:relatedwork}

To the best of our knowledge, ours is the first work to: (1) Present techniques on finding blocking functions by explicitly trading-off recall for efficiency, and in a more expressive tree-based structure than flat conjunctive structures of past work; (2) Formally introduce and study the problem of {\em rollup} as an important post-processing step to assemble small canopies and increase recall; (3) Provide automatic solutions to the {\em drill-down} problem as a way of bootstrapping blocking with no manual effort, or augmenting manually-generated hash functions; (4) Present an automatic blocking system for de-duplication in a distributed setting that is applied to two large commercial datasets from a search engine. Very few pieces of previous work consider blocking based on labeled training data, while there is a much larger body of work on hand-tuned blocking techniques using similarity functions. We start by describing the relationship of our work with blocking based on labeled data (Section~\ref{subsec:rel2}), followed by blocking without labeled data (Section~\ref{subsec:rel1}), and finally other work on de-duplication (Section~\ref{subsec:relother}).


\subsection{Blocking With Labeled Data}
\label{subsec:rel2}

Two recent papers~\cite{bilenko06:blocking,knoblock} presented approaches to constructing a blocking function using a labeled dataset of positive and negative examples. Roughly speaking, both papers learn conjunctive rules (and disjunctions of conjunctive rules) to maximize recall.~\cite{bilenko06:blocking} attempts to maximize the number of positive minus negative examples covered, effectively using negative examples as a proxy for minimizing the size.~\cite{knoblock} uses only positive examples, but does not explicitly incorporate any size restriction. Below we give a detailed comparison with these past approaches: 


\squishlist
\item We present \btree s, a more expressive language for expressing disjoint
blocking functions than previous work. Given only simple or conjunctive
blocking functions, it may not be possible to construct an effective blocking
function without a large number of map-reduce rounds (disjuncts).  
\eat{This is particularly
true if some attributes are missing (null) for a significant fraction of
entities. Even if blocking functions are extended to allow exclusion of null
values, no true positives where one entity has a null for some attribute $a$
can be covered by a blocking function that involves $a$.  However, since each
round (each disjunct) leads to a separate map-reduce job, it is important to
limit solutions to as few disjuncts as possible.}

\item Minimizing negative training examples covered by a blocking solution may lead to quality problems from {\em overly aggressive blocking}.   For example, consider a movie and a remake with the same title but released in a different year -- while the two are a negative example, this does not mean it is a bad idea to block movies together when their titles are similar.  In short, blocking should be optimized for {\em recall} vs. {\em efficiency}, and match rules optimized for {\em precision}. 

\item Minimizing negative training examples does not match the cost model of
{\em parallel computation} models like map-reduce, where latency is determined
by the {\em largest block}.  
\eat{In our setting, as
in~\cite{vernica10}, matching is performed at scale on a map-reduce
infrastructure.  In this setting,  the total latency is determined by the size
of the {\em largest block}, and a job will fail if a single block exceeds the
available memory on a single machine.}

\item We are the first to introduce and solve the rollup and drill down problems. These problems were not addressed in~\cite{bilenko06:blocking,knoblock}, or in other past work on blocking.


\squishend

\subsection{Blocking Without Labeled Data}
\label{subsec:rel1}

\cite{merge-purge} introduced the notion of blocking (called ``merge-purge'')
by constructing a key for each record, sorting based on the key, and then
performing matching and merging in a sliding window.~\cite{merge-purge} (and
other variants~\cite{kelley85,jaro89}) do not consider automatic generation of
optimal blocking functions in a distributed environment, based on training
data.
\eat{did
not consider blocking in a distributed environment (for instance, making a
sort too expensive); moreover, automatic generation of optimal blocking
functions based on training data was not considered in~\cite{merge-purge}.
Other variants~\cite{kelley85,jaro89} of such an approach have been proposed
earlier, for instance where blocks consist of exact key matches.}

SWOOSH~\cite{swoosh} is a recently developed generic entity resolution system
from Stanford. Their specific paper on blocking~\cite{swoosh-blocking} focuses
on ``inter-block communication'', by propagating matched records to other
blocks. Once again, automatic generation of blocking functions is not the
subject of~\cite{swoosh-blocking}. Further, D-Swoosh~\cite{d-swoosh} (and
other similar work~\cite{plink, thetajoin}), their distributed framework for
entity resolution focus on distributing pairwise comparisons across multiple
processors, as opposed to our focus of partitioning the data to reduce the
number of total pairwise comparisons.

Reference~\cite{canopy-kdd} presented techniques for generating non-disjoint
canopies based on distance measures such as jaccard similarity of tokens.
After choosing a distance function, they pick records as canopy centers, and
add to each canopy all records that are within some distance based on the
distance measure.\eat{ (All records that are very close to currently chosen canopy
centers are removed from future consideration as centers.)} The algorithms
from~\cite{canopy-kdd} cannot be directly scaled to a distributed environment.
A similar approach of generating (non-disjoint) canopies by clustering based
on any distance measure was also proposed in~\cite{rendle08}. Some other
work~\cite{chch03} considers blocking based on bi-grams  of string attributes,
followed by creation of inverted lists for each bigram. Another recent piece
of work~\cite{jlm03} considered transforming the data into a euclidean space.
While the above approaches weren't designed specifically for a distributed
environment, recently~\cite{vernica10} studied the problem of performing
approximate set similarity joins using a map-reduce framework. Their work can
be used for blocking when records are compared for duplicates based on set
similarity functions. Also, a recent system, MAHOUT~\cite{mahout}, described
an implementation of canopy clustering in a map-reduce framework. Finally,
~\cite{baxter} performed a comparative study of blocking strategies
from~\cite{merge-purge,canopy-kdd,chch03,jaro89}.

In general, the approaches described above rely on the knowledge of specific similarity/distance functions. Furthermore, they necessarily generate non-disjoint canopies, whereas one primary goal of our work was to consider disjoint canopies as an important choice for distributed de-duplication and obtain non-disjointness as multiple rounds of disjoint sets of canopies. Finally, none of this past work considers the rollup and drill-down problems.

\subsection{Other Work}
\label{subsec:relother}

De-duplication has been studied for over 50 years now, starting with the seminal pieces of work in~\cite{er59,fsunter69}. De-duplication of very large datasets broadly proceeds by performing blocking, followed by pair-wise (or cluster-wide) similarity computation within each block. A large body of work has focused on the latter step of pair-wise similarity computation, known as matching~\cite{koudas06, secondstring,fsunter69}. Some other work~\cite{cfuzzy,kfuzzy,gfuzzy} has considered fuzzy matching in the context of databases, however none of this work considers the problem of automatic blocking, drill-down, or rollup. Finally, we note that the structure of \btree s is akin to that of decision trees, a popular approach to classification; however, we note that the objectives of our \btree s are completely different, that of effectively trading off recall for efficiency in deduplication. 


\section{Preliminaries}
\label{sec:components}

\subsection{Background and Notation}

We use $U$ to denote the set of entities (i.e., records) to be de-duplicated. Dividing $U$ for pairwise comparisons is known as {\em blocking} (or {\em canopy formation}). The divided pieces are called {\em blocks} (or {\em canopies}). We use ${\cal C}$ to denote the set of canopies, and $C_i$'s denote the individual canopies. Formally, given a universe $U$, a set of canopies is given by a finite collect ${\cal C}=\{C_1, \ldots, C_k\}$, $C_i\subseteq U$ and $\bigcup_i C_i = U$. A specific method to construct ${\cal C}$ from $U$ is called a {\em blocking function}. We start by restricting our attention to blocking functions that create a {\em disjoint} set of canopies (i.e., if $i\neq j$, then $(C_i\cap C_j)=\emptyset$) and then extend our results for {\em non-disjoint} sets of canopies (Section~\ref{sec:nondisjoint}). Intuitively, a good blocking function must satisfy two desirable properties. First, canopy formation increases the efficiency of de-duplication by eliminating the need for performing pairwise comparisons between all pairs of entities in $U$. Second, the quality of de-duplication (i.e., recall of identified duplicates) must not be significantly reduced by performing fewer comparisons. Therefore, our goal is to find a set of canopies such that most duplicates in $U$ fall within some canopy. We shall use ${\cal T}^+\subset U\times U$ to denote a training dataset consisting of labeled duplicates in $U$, over which recall of blocking functions is measured. We shall construct blocking functions using a space ${\cal H}$ of {\em hash functions} that partition $U$ based on attributes of the entities in $U$; each hash function assigns one hash value for each entity. For example, one hash function partitions $U$ based on the first character of the titles of movies. A {\em conjunction} of hash functions $h_1, \ldots, h_l$ is equivalent to creating a single hash value by concatenating the hash values obtained by each $h_i$, effectively creating partitions (equivalence classes) where values of each of the hash functions matches. Typically ${\cal H}$ is generated manually based on domain knowledge, and we shall present techniques to construct blocking functions using any ${\cal H}$. In addition, we shall also present techniques to automatically identify optimal hash functions for each attribute (Section~\ref{sec:hashsearch}).

\subsection{Cost Model}
\label{subsec:cost}

While \cobbler\ can be configured with any {\em cost model} for optimizing canopy formation, we use {\em latency} as the default cost model in our discussion.\footnote{All our algorithms and complexity results carry over for any ``monotonic cost function'', i.e., $\mbox{cost}({\cal C})\leq \mbox{cost}({\cal C}')$ whenever $\forall C\in {\cal C}, \exists C'\in {\cal C}'$ such that $C\subseteq C'$.} The latency of any canopy formation is given by the total time it takes to perform all pairwise comparisons in each canopy.

In a grid environment (such as our de-duplication system implemented using map-reduce), pairwise comparisons in the set of canopies are performed in parallel. Given a canopy formation  ${\cal C} = \{C_1, \ldots, C_k\}$, with number of entities in canopy $C_i$ denoted by $s_i$, the total number of pairwise comparisons being performed is $\sum_{i=1}^k s_i^2$. Motivated by de-duplication in a grid environment, we use the cost model $\mbox{cost}({\cal C})=\max_i{s_i}$: Clearly, in a truly elastic grid with a potentially infinite supply of machines, pairwise comparisons for each canopy are performed on a separate machine. Therefore, the latency is given by the largest canopy, justifying our cost model of using $\max_i{s_i}$.

When the number of machines on the grid are limited (and specifically, when there are fewer machines than canopies), we are faced with the problem of assigning canopies to machines. The following theorem shows that this assignment is NP-hard in general, based on a direct reduction from a scheduling problem. However, we also show that the latency using the largest canopy gives an upper bound on the best possible assignment.


\begin{theorem}
Given a set ${\cal M}=\{M_1, \ldots, M_m\}$ of $m$ machines, a canopy formation ${\cal C} = \{C_1, \ldots, C_k\}$ over $N$ entities, $m<k$, any assignment $A:{\cal C}\rightarrow M$ of canopies to machines has a cost given by $\mbox{cost}_A({\cal C})=\max_{j=1}^m (\sum_{C_i: A(C_i)=M_j} |C_i|^2)$. We have:
\squishlist
\item[1.] It is NP-hard to find an assignment that minimizes  $\mbox{cost}_A({\cal C})$.
\item[2.] \sloppy For all assignments $A$, we have $\max_{i=1}^k{|C_i|^2} \leq \mbox{cost}_A({\cal C})\leq (1+\frac{k}{m})\max_{i=1}^k{|C_i|^2}$. More specifically, let $X=\max{(\max_{i=1}^k{|C_i|^2},\frac{\sum_{i=1}^k{|C_i|^2}}{m})}$. We have $X\leq \mbox{cost}_A({\cal C})\leq 2X$.
\squishend
\end{theorem}
\eat{
\begin{proof} (Sketch.)
\ \newline
\noindent {\bf (1) NP-hardness} directly follows from the well-known {\em minimum makespan scheduling} problem~\cite{garey}: Given $n$ jobs with processing times $p_1, \ldots, p_n$, it is NP-hard to find  an assignment of jobs to $m$ identical machines to minimize the completion time.

\noindent {\bf (2) Approximation:} Since all pairwise comparisons for the largest canopy must happen on some machine, and since a perfect assignment would equally split all pairwise comparisons, we have $X\leq \mbox{cost}_A({\cal C})$. Next we obtain an upper bound. Consider an optimal assignment $A^*$, and let $\mbox{cost}^p$ denote the total latency on machine $p$. Note that in the optimal assignment, no specific canopy can be moved out of the most overloaded machine to the least overloaded machine to reduce the latency. Therefore, the difference in latency between any two machines must be at most the latency due to the largest canopy. Therefore, we have:
$$\mbox{cost}_A({\cal C}) \leq \mbox{cost}^p + \max_{i=1}^k{|C_i|^2}$$
\noindent By summing the equation above for all $p\leq m$, we have:
$$\mbox{cost}_A({\cal C}) \leq \frac{\sum_{p=1}^m \mbox{cost}^p}{m} + \max_{i=1}^k{|C_i|^2}$$
$$ = \max_{i=1}^k{|C_i|^2}+\frac{\sum_{i=1}^k{|C_i|^2}}{m} \leq 2X$$
Also, using $|C_i|\leq  \max_{i=1}^k{|C_i|}$ in the above equation, we have $\mbox{cost}_A({\cal C}) \leq (1+\frac{k}{m})\max_{i=1}^k{|C_i|^2}$.
\end{proof}
}

\noindent Based on the theorem above, henceforth, we focus on the problem of finding best canopies that satisfy the constraint of $\max_i{s_i}\leq S$, for some given $S$.

\eat{
\subsection{Outline of the rest of the paper}

[[Change this part]]
The following common problems that need to be addressed for each canopy-formation method are the topic of the next three sections:

\begin{itemize}
\item Canopy Formation Algorithms (Section~\ref{sec:core}): Core contribution of solving the optimization problem, given a space of blocking predicates, and some techniques for estimating canopy sizes for combinations of predicates (or single predicates?)
\item Estimating the sizes of canopies (Section~\ref{sec:estimation}): Sub-problem arising in Step-(1) above: Given a specific blocking predicate (or a combination of blocking predicates), how do we estimate the size of each canopy, based on each canopy formation method.
\item Block Predicates (Section~\ref{sec:space}): What is a good space of blocking predicates to use? Given a dataset like the one Ankur is creating, what all blocking functions do we use? The main question here is what would be a principled way to obtain a search space (single-attribute and combinations of attributes, and various functions on each attributes)
\end{itemize}
}

\section{Blocking Based on Labeled Data}
\label{sec:labeled}

This section addresses the problem of constructing disjoint blocking functions using a labeled dataset of positive examples. After formally defining the problem (Section~\ref{subsec:formulation}), we introduce a tree-structured language for expressing blocking functions (Section~\ref{sec:btree}). We then show that the general problem of finding an optimal blocking function is NP-hard (Section~\ref{subsec:btreehard}), and finally we present a greedy heuristic algorithm (Section~\ref{subsec:btreegreedy}) to find an approximate blocking function.

\subsection{Problem Formulation}
\label{subsec:formulation}

We formally define the problem of creating canopies given labeled data consisting of examples of duplicates (positive pairs). Recall the two conflicting goals of canopy formation: The more divisive a set of canopies is, the more likely it is to miss out on true duplicates. We formulate an optimization problem that trades off the two objectives of canopy formation, by associating a hard constraint on the maximum size of each canopy and maximizing the number of covered positive examples (recall) subject to this size constraint.

\begin{definition}[Blocking Problem]\label{defn:labeled}
Given a labeled set ${\cal T}^+$ of positive examples, a space ${\cal H}$ of hash functions, a size bound $S$ on every canopy, and a size function $size()$ that returns the size of a canopy obtained by applying any conjunction of hash functions in ${\cal H}$ on any input dataset $I$, construct a disjoint blocking function ${\cal B}$ that partitions any input $I$ into a set ${\cal C}$ of disjoint canopies of size at most $S$, while maximizing the number of pairs from ${\cal T}^+$ that lie within canopies, i.e., maximizing: $recall = \frac{|\{(r_1,r_2)\in {\cal T}^+ | \exists c\in C, r_1,r_2\in C\}|}{|{\cal T}^+|}$.
\end{definition}

\noindent We make a few important observations about our problem definition. (1) As a reminder, we start by considering only disjoint blocking, and extend to non-disjoint blocking in Section~\ref{sec:nondisjoint}. The next section describes a language to represent disjoint blocking functions (${\cal B}$), and subsequently we give algorithms for finding ${\cal B}$. (2) We assume that there is a known size estimation function. In practice, some previous work on blocking~\cite{bilenko06:blocking} has used negative examples as an indirect way of incorporating size restrictions. Alternatively, previous work on estimating the cardinality of selection queries using histograms (refer~\cite{ioannidis03}) can be used to estimate canopy sizes, as we shall see each canopy is obtained as a conjunction of hash functions. Of course, if the entire dataset were available during the construction of blocking predicates, it could be used for size computation. (In particular, exact size computation for the blocking technique we propose can be done in a few scans. Also, we shall see that our technique can be adaptively applied even in case of inaccurate size estimates.) (3) For this section we assume the existence of a space ${\cal H}$ of hash functions. Most previous work has assumed the manual creation of such atomic hash functions. We also present in Section~\ref{sec:hashsearch} an automated method of enumerating hash functions for each attribute. (4) Finally, we assume the positive examples ${\cal T}^+$ are known; we describe the construction of this dataset in the experiments section (Section~\ref{sec:experiments}).

\subsection{Blocking-Tree Space}
\label{sec:btree}
This section presents a generic language for expressing disjoint blocking functions. We introduce a {\em hierarchical blocking tree} (called {\em \btree}), that partitions the entire set of entities in a hierarchical fashion by successively applying atomic hash functions from a known class ${\cal H}$. Formally:

\begin{definition}
A \btree\ $B=(N,E,h)$ is composed of a tree with nodes $N$ and edges $E$, and $h:N\rightarrow {\cal H}$ maps each node in the tree to a particular hash partitioning function from ${\cal H}$.
\end{definition}

\noindent Intuitively, each leaf node of the tree corresponds to a canopy. The \btree\ is built using the inputs described in Definition~\ref{defn:labeled}, namely the training data, a known space of atomic hash functions ${\cal H}$, and canopy-size estimates. Each node $n\in N$ in the tree corresponds to a set of entities from the entire set obtained by applying the hash functions from the root down to $n$. Each node $n$ (with a size estimate exceeding the allowed maximum) then applies a particular partitioning hash function to create disjoint partitions of the set of entities corresponding to $n$. 

At run-time, each entity is run through the \btree, and directed to the machine in the cluster based on the leaf node. (Note that in a distributed environment, the entire data itself is initially partitioned across multiple machines; therefore, the \btree\ is stored on every machine in order to redistribute the data based on the canopies.)  Note that in practice the total number of large canopies created by any hash function on any node is a constant, for instance due to NULL values in the data, or a common default value for an attribute. Therefore, the size of the constructed \btree\ in terms of the number of nodes is small, so that the \btree\ fits in memory, and applying the \btree\ to an entity is efficient.

\begin{figure}[t]
\centering
\includegraphics[width=2.2in]{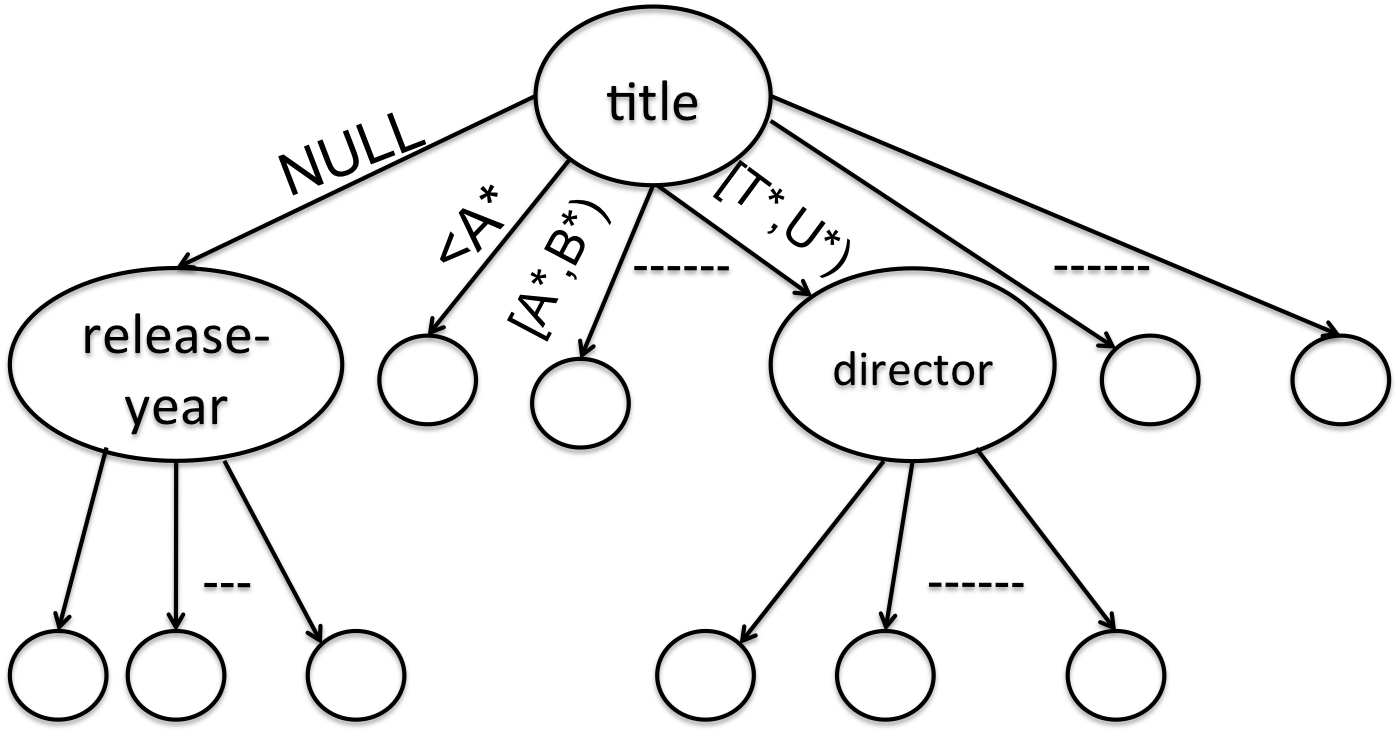}
\vspace{-3mm}
\caption{\label{fig:btree}\small Example of a tree-structured disjoint blocking function.}
\end{figure}

\begin{example}
Figure~\ref{fig:btree} shows an example \btree\ for movie data with the root partitioning the movies lexicographically based on the title. This partition results in two large canopies---the node corresponding to {\tt NULL} titles, and the node corresponding to titles that start with ``T'' (assume all titles have been capitalized in advance). In the {\tt NULL} canopy a partition based on the release-year of the movie is performed, while the movies starting with ``T'' are partitioned by the name of the movie's director. All leaf nodes in the resulting tree satisfy the maximum canopy-size requirement, and hence no further partitioning is performed.
\end{example}

\subsubsection{Restricted languages}
\label{sec:restricted}
We note that \btree s are a very expressive language to describe disjoint blocking functions. In particular, the following natural languages are obtained as restrictions of \btree s:

\squishlist
\item[1.] {\bf Single hashes:} Clearly single hash functions are equivalent to a \btree\ of height $1$.
\item[2.] {\bf Conjunctive functions (chains):} Conjunctions of hash functions are equivalent to restricting the width of \btree s to $1$, i.e., a branching factor of $1$. In particular, a conjunction $h_1\wedge \ldots \wedge h_k$ is equivalent to applying each hash function $h_i$ in sequence to every single canopy, irrespective of whether a canopy is smaller than the required size $S$. Note that (disjunctions of) chains are the basic construct used in~\cite{bilenko06:blocking,knoblock}, where as our language of (disjunctions of) \btree s is significantly more expressive.
\item[3.] {\bf Chain-tree:} {\em Chain-trees} are an extension of conjunctions where we are again specified a chain $h_1, \ldots, h_k$ of hash functions to be applied in sequence, however, subsequent hash functions are applied only if the canopy size exceeds the allowed maximum. In particular, chain-trees are obtained by restricting every level of \btree s to have the same hash function.
\squishend

\noindent In our experiments, we implement algorithms for \btree s and all the restricted languages above and compare them in terms of recall, to observe a significantly higher recall using \btree s.

\subsection{Intractability}
\label{subsec:btreehard}

Next we demonstrate that the general problem of finding the optimal \btree\ is NP-hard, and subsequently present a heuristic greedy algorithm.

\begin{lemma}[\btree\ intractability]
Given a training set ${\cal T}^+$ with positive examples, a space ${\cal H}$ of hash functions, and a bound $S$ on the maximum size of any canopy, assuming $P\neq NP$, there does not exist any polynomial-time (in ${\cal T}^+,{\cal H}$) algorithm to find the optimal \btree.
\end{lemma}
\subsection{Greedy Algorithm}
\label{subsec:btreegreedy}

\begin{algorithm}[t]
{\scriptsize
\caption{Recursive greedy construction of \btree.}
\begin{algorithmic}[1]
\STATE {\bf Input:} Node $n$ consisting of entities $C_n$, duplicates ${\cal T}^+$, space ${\cal H}$ of hash functions, size bound $S$.
\IF{$|C_n|>S$}
\STATE $least=\infty$; $best=${\tt NULL}
\FOR{$h\in {\cal H}$}
\STATE Compute $e=\mbox{elim-count}(C_n,S,h)$
\IF{$least>e$}
\STATE $least=e$; $best=h$
\ENDIF
\ENDFOR
\STATE Set $best$ as the hash for node $n$.
\STATE Recurse on nodes resulting from $best$ applied to $n$.
\ENDIF
\end{algorithmic}
\label{algo:btree}
}
\end{algorithm}

We propose a simple heuristic for constructing the \btree\ described in Algorithm~\ref{algo:btree}. The general scheme of the algorithm is to locally pick the best hash function at every node in the tree, if the size (estimate) of the number of entities in this node is over the allowed maximum $S$. (If a particular hash function generates many large canopies, it is ignored, in order to maintain a small \btree. However, as described before, the number of large canopies is typically small; in our experiments over 140K movie entities, no hash function created more than a few large canopies.) The best hash function for a node is picked greedily by counting for all hash functions $h\in {\cal H}$, the number of duplicates that get eliminated on choosing the hash function $h$. The hash function that minimizes the number of eliminated duplicates is chosen. We describe three ways of counting the number of examples eliminated (function {\em elim-count} in Algorithm~\ref{algo:btree}). Suppose a node $n$ has $P_n$ positive pairs, and application of $h$ eliminates $P_h$ duplicates and creates canopies $C_1, \ldots, C_k$ exceeding size $S$ (among other canopies that are smaller than $S$). If the number of positive pairs in $C_i$ is denoted $P(C_i)$, then the three ways of counting the drop in the number of positive pairs are as follows:

\begin{itemize}
\item {\bf Optimistic Count:} Intuitively, Algorithm~\ref{algo:btree} picks the hash function by assuming that no more duplicate examples would get eliminated, hence it is optimistic:
$$\mbox{Optimistic}=P_h$$

\item {\bf Pessimistic Count:} On application of a hash function $h$, we say that the number of duplicates that are eliminated include the ones broken by $h$ as well as all examples that still remain in canopies larger than $S$:
$$\mbox{Pessimistic}=(P_h+\sum_{i=1..k}P(C_i))$$

\item {\bf Expected Count:} For the duplicates that still remain in large canopies after applying $h$, we compute an expected number of eliminated duplicates based on a random split of the canopy so as to obtain canopies of size $S$:
$$\mbox{Expected}=(P_h+\sum_{i=1..k}\frac{P(C_i)(n_i-1)}{n_i})$$
\noindent where $n_i=\lceil \frac{|C_i|}{S} \rceil$.  Effectively, a random split would only retain a $\frac{1}{n_i}$ fraction of the positive pairs, assuming each pair is independent.
\end{itemize}

Finally we note that an important feature of constructing the \btree\ is that it can be naturally adapted at runtime based on the actual canopy sizes, such as when the canopy-size estimates turned out to be inaccurate, or when available memory has reduced. Suppose while construction of the \btree\ a canopy-size bound of $S$ (say, 5000 entities) was imposed, we may choose to construct the \btree\ based on a maximum canopy-size of a fraction of $S$ (say, 1000 entities). Effectively, we will create a longer tree than necessary, and this ``extra'' portion of the tree may be used if any canopy needs to be split further based on the reasons described above. Conversely, if the actual canopy sizes turn out to be smaller than expected, we may choose to run through only a smaller part of the tree.

\section{Rolling up small canopies}
\label{sec:rollup}

In this section, we introduce the problem of rolling up small canopies. The primary motivation for studying this problem is that a blocking function may unnecessarily have to create many small canopies, in order to make some of the larger canopies fit the required size bound. Therefore, as a post processing step, we can take the result of any blocking function, and combine multiple small canopies maintaining the size requirement yet increasing the overall recall.

We are given a set of canopies ${\cal C} = \{C_1, C_2, \ldots, C_m\}$, where each canopy $C_i$ has size (much) less than our canopy size limit $S$. We are also given a set of pairs of matching records ${\cal T}^+ = \{\ldots, (r_{i_1}, r_{i_2}), \ldots \}$. The rollup problem is to find a set of canopies ${\cal D} = \{D_1, D_2, \ldots, D_\ell\}$ such that
\squishlist
\item {\em Disjointness Constraint}: $\forall i, j$, $i\neq j$, $D_i \cap D_j = \emptyset$
\item {\em Roll Up Constraint}: $\forall i$, $\exists i_1, i_2, \ldots$, such that, $D_i = \cup_j C_{i_j}$
\item {\em Maximum Size Constraint}: $\forall i,\ |D_i| \leq  S$
\item {\em Maximize Recall}: minimize the number of pairs of matching examples from ${\cal T}^+$ that are split across canopies.
\squishend

\noindent Note that the rollup problem can be applied on any set of canopies
generated using any previous blocking function. In particular, it can be
applied on the \btree\ blocking function generated in
Section~\ref{sec:labeled}. Each leaf of the \btree\ corresponds to a canopy,
and by applying rollup, some leaves of the \btree\ get merged so as to
maintain the size requirement but increase recall. Figure~\ref{fig:rollup}
shows an example blocking function obtained by performing rollup on the
\btree\ in Figure~\ref{fig:btree}. Note that although the resulting blocking
function isn't a tree, the resulting DAG can still be used for distributed
canopy formation: Each entity starts at the root and traverses all the way
down through the directed edges to a (possibly rolled-up) leaf node, which
corresponds to a canopy. 

We start by showing that the roll-up problem is intractable:

\begin{figure}[t]
\centering
\includegraphics[width=2.2in]{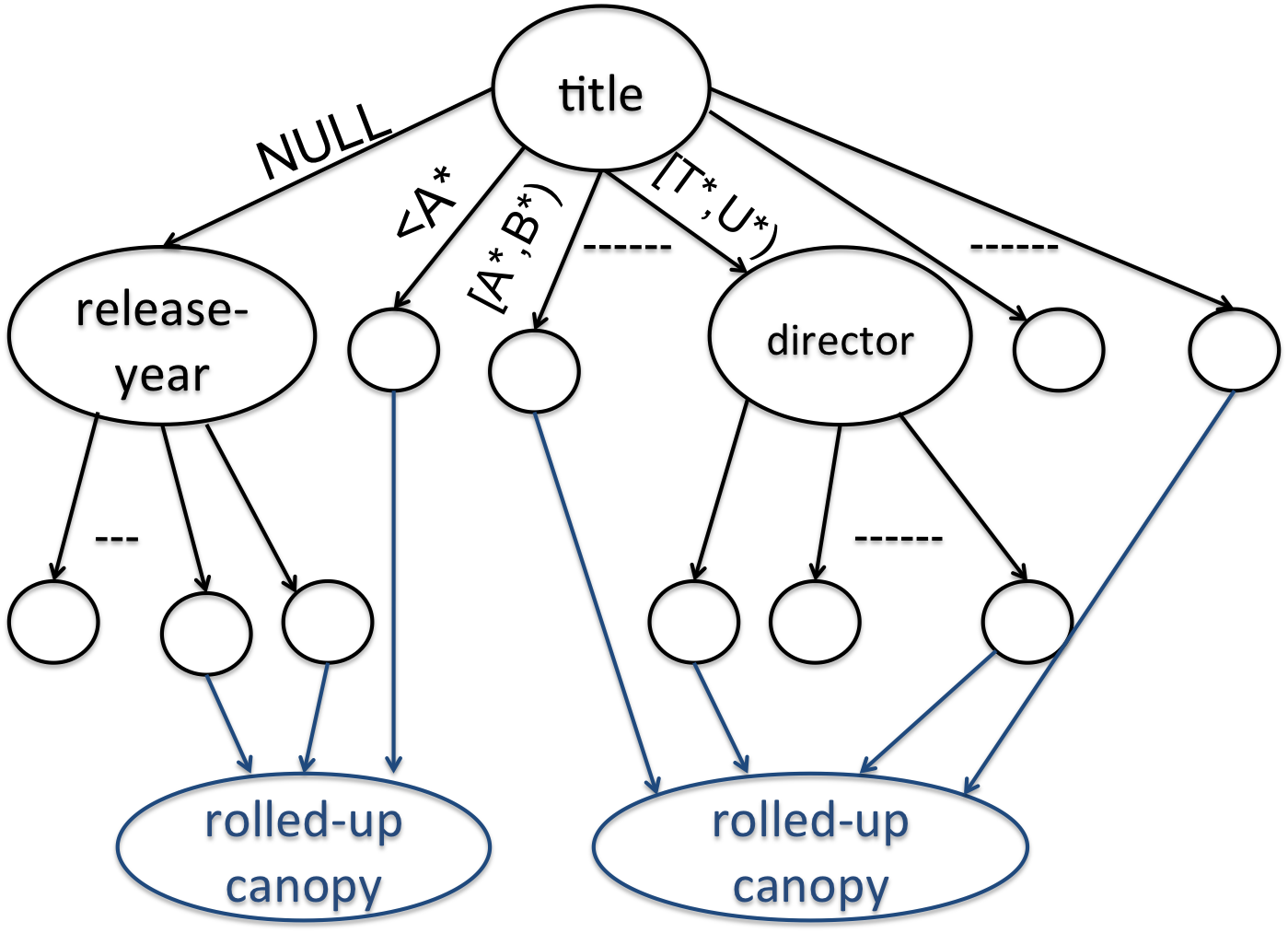}
\caption{\label{fig:rollup}\small Rollup applied on canopies (leaf-nodes)
generated in Fig.~\ref{fig:btree}.}
\end{figure}


\begin{lemma}[NP-completeness]
The rollup problem described above is NP-complete.
\end{lemma}
\eat{
\begin{proof}(sketch)
The problem is clearly in NP. Given a solution, which is polynomial in the size of the input, one can verify in PTIME that all the constraints are satisfied.

To show NP-hardness, we reduce the $0/1$-Knapsack problem to the rollup problem. Suppose $B = \{b_1, b_2, \ldots, b_m\}$ be the set of items from the knapsack instance. Let $k$ be the capacity of the knapsack. Denote by $w(b_i)$ and $p(b_i)$ the weight and profit, resp., of each item. Without loss of generality assume that $\forall i$, $p(b_i) < w(b_i)$ (we will need this in the reduction); otherwise multiply the weights and $k$ by a constant such that the weights are smaller than the profit.

We reduce this to an instance of the rollup problem as follows. We create one set of records $C_i$ for each $b_i$ and a ``sink'' set $C_0$. We create $q = \sum_i p(b_i)$ pairs of matching examples; for each $i$, we place $p(b_i)$ records in $C_i$ and their corresponding matching records in $C_0$. We fill in distinct records into the $C_i$ sets to make their size equal $w(b_i)$. Let the maximum size limit for any canopy in the solution be $q + k$. Any solution for this structure would assign sets $\{C_{i_1}, C_{i_2}, \ldots\}$ to the same canopy as $C_0$; since every matching example pair has one record in $C_0$, the number of pairs of matching examples that are split across canopies would be
\[\sum_i p(b_i) - \sum_j p(b_{i_j})\]
It is easy to see that there is a knapsack solution with profit $p$ if and only if there is a rollup solution with recall $q-p$.
\end{proof}
}

\renewcommand{\setminus}{\ensuremath{-}}
\renewcommand{\mid}{\ensuremath{\ |\ }}

\begin{algorithm}[t]
{\scriptsize
\caption{Greedy Canopy Rollup Algorithm}
\begin{algorithmic}[1]
\STATE {\bf Input:} ${\cal C} = \{C_1, C_2, \ldots, C_m\}$, set of matching pairs ${\cal T}^+$, maximum canopy size $k$
\STATE Set ${\cal D} \leftarrow {\cal C}$ {\em // initialize}
\REPEAT
\STATE {\em // Candidate pairs that can be merged}
\STATE ${\cal D}_{pair} \leftarrow \{(D_1, D_2) \mid |D_1| + |D_2| \leq k\}$
\IF{${\cal D}_{pair} \neq \emptyset$}
\STATE $(D_1^\star, D_2^\star) \leftarrow \arg\max_{{\cal D}_{pair}} \frac{benefit(D_1, D_2)}{\min(|D_1|, |D_2|)}$
\STATE {\em // Merge $D_1^\star$ and $D_2^\star$ into one canopy}
\STATE ${\cal D} \leftarrow {\cal D} \cup \{D_1^\star \cup D_2^\star\}\setminus \{D_1^\star, D_2^\star\}$
\ENDIF
\UNTIL{${\cal D}_{pair} = \emptyset$}
\STATE {\bf Return} ${\cal D}$
\end{algorithmic}
\label{alg:rollup}
}
\end{algorithm}

\eat{
\subsection{Greedy Rollup Algorithm}
We used a reduction from the knapsack problem to show NP-hardness of the rollup problem. However, this reduction is only one way; one can not use a solution to an appropriately constructed knapsack instance to solve the rollup problem. Hence, while the knapsack problem has pseudo-polynomial and 2-approximate algorithms, they cannot be directly applied to obtain approximation guarantees on the rollup problem.
}

Next we propose a greedy heuristic for the rollup problem that is inspired by Dantzig's 2-approximation algorithm \cite{dantzig57:greedy} for the knapsack problem. Conceptually, our algorithm (Algorithm~\ref{alg:rollup}) starts with the initial set of canopies, and progresses in steps. In each step, the algorithm finds the pair of sets $D_1, D_2$ that together have less than $S$ records, and maximize the following quantity:
\begin{equation}
benefit(D_1, D_2) / \min(|D_1|, |D_2|)
\end{equation}
$benefit(D_1, D_2)$ is the number of matching pairs $(r_{i_1}, r_{i_2}) \in P$ such that $r_{i_1} \in D_1$ and $r_{i_2} \in D_2$. Intuitively, in each step we pick the canopy that has the smallest size but also puts a large number of matching pairs in the same canopy.

Algorithm~\ref{alg:rollup} can be efficiently implemented in time linear in the number of matching pairs ($|{\cal T}^+|$) and quadratic in the number of input canopies ($|{\cal C}|$). Initially, we compute for each canopy $D \in {\cal C}$, one {\em merge candidate}. This is a canopy $D'$ such that $|D'| \geq |D|$ and $|D| + |D'| \leq S$ such that $benefit(D, D')$ is maximum. This step takes $O(|{\cal T}^+|\cdot |{\cal C}|^2)$ time. In each step, we find the canopy $D$ whose merge candidate has the maximum benefit to size ratio; we then merge $D$ with its merge candidate. The new merge candidate for a canopy other than $D$ and $D'$ is either $D \cup D'$ or its old merge candidate -- this step takes $O(1)$ time for each canopy. The new merge candidate for $D \cup D'$ can be computed in $O(|{\cal T}^+|\cdot |{\cal C}|)$ time by considering all the other canopies and the positive examples. Since the algorithm terminates in at most $|{\cal C}|$ steps, our algorithm has $O(|{\cal T}^+|\cdot |{\cal C}|^2)$ time complexity.

\section{Drill-Down Problem}
\label{sec:hashsearch}

In Section~\ref{sec:labeled} we assumed a pre-existing and manually-generated space of hash functions (as is done in most previous work).  Next we propose automatic (only using an attribute's domain and labeled dataset) techniques for generating hash functions. Automatically constructed hash functions may be used to bootstrap the blocking methods, eliminating the need for a significant upfront manual effort. Moreover, even in the presence of an existing space of manually constructed hash functions, we can augment the space with (better) automatically generated hash functions.

We introduce the ``drill-down'' problem for a single attribute. Our goal is to optimally divide a single-attribute's domain into disjoint sets so as to cover as many duplicate pairs as possible, but ensuring that the cost associated with any set is below a required threshold. First we formally define the partitioning of an attribute's domain into {\em disjoint, covering, contiguous} subsets (called a {\em DCC partition}), then define the problem of finding an optimal DCC partition.

\begin{definition}[DCC Partition]\label{defn:dcc}
Given a domain $D$ with total ordering $\prec$, least element `$start$' and greatest element `$end$'\footnote{The least and greatest element may be part of $D$ in some cases (e.g., all 10-digit phone numbers) and not part of $D$ in others (e.g., $-\infty$ and $+\infty$ for real numbers).}, we say that a set ${\cal I}$ is a DCC partition of $D$ if $\forall I\in {\cal I}: I\subseteq D$ and all of the following hold:
\squishlist
\item {\bf Disjoint:} $I_1,I_2\in {\cal I}, I_1\neq I_2 \Rightarrow I_1\cap I_2=\emptyset$
\item {\bf Contiguous subset:} Every $I\in {\cal I}$ is of the form $[I^1,I^2]$, $[I^1,I^2)$, $(I^1,I^2]$, or $(I^1,I^2)$, $I^1\preceq I^2$ and $I^1,I^2\in D_A\cup\{start,end\}$
\item {\bf Covering:} $D_A = \bigcup_{I\in {\cal I}} I$ \rbox
\squishend
\end{definition}

\noindent Intuitively, a DCC partition completely divides $D$ by ``tiling'' the entire domain. Also, note that the total ordering doesn't need to correspond to the ``natural ordering'' such as lexicographic for strings or `$<$' ordering for numeric. For instance, may choose to order director names by their last name and find a hash function, then also order them by first name and find another hash function.

Next we formally define the drill down problem.

\begin{definition}[Drill Down Problem]
Consider a single attribute $A$ with an ordered domain
$D_A,\prec,start,end$, a set of $n$ duplication pairs ${\cal T}^+ =
\{(a_1^1,a_1^2), \ldots, (a_n^1,a_n^2)\}$, $\forall i: a_i^j\in D_A,
a_i^1\prec a_i^2$, and any monotonic black-box cost
function\footnote{$\forall I,I'\subseteq D_A: I\subseteq I' \Rightarrow
\mbox{cost}(I)\leq \mbox{cost}(I')$. Note that, in practice, for uniformly
distributed data the cost function may simply bound the total size of the
interval. But for skewed data, the size of the interval depends on the density
of the data; therefore, we allow any arbitrary cost function.}
$\mbox{cost}:I\subseteq D\rightarrow \mathbb{R}$, and a maximum cost bound $S$
on any partition. Our goal is to find a DCC partition ${\cal I}$ of $D_A$ such
that:
\squishlist
\item[1.] $\forall I\in {\cal I}: \mbox{cost}(I)\leq S$
\item[2.] Let $\mbox{cov}({\cal I},{\cal T}^+)$ be the number of duplicates
covered: $\mbox{cov}({\cal I},{\cal T}^+) = \sum_{1\leq i\leq n: \exists I\in
{\cal I} \mbox{ with } [a_i^1,a_i^2]\subseteq I} 1$. For
any DCC partition ${\cal I}'$ satisfying (1) above,  $\mbox{cov}({\cal
I},{\cal T}^+)\geq  \mbox{cov}({\cal I}',{\cal T}^+)$.\rbox
\squishend
\end{definition}

\begin{figure}[t]
\centering
\includegraphics[width=2.2in]{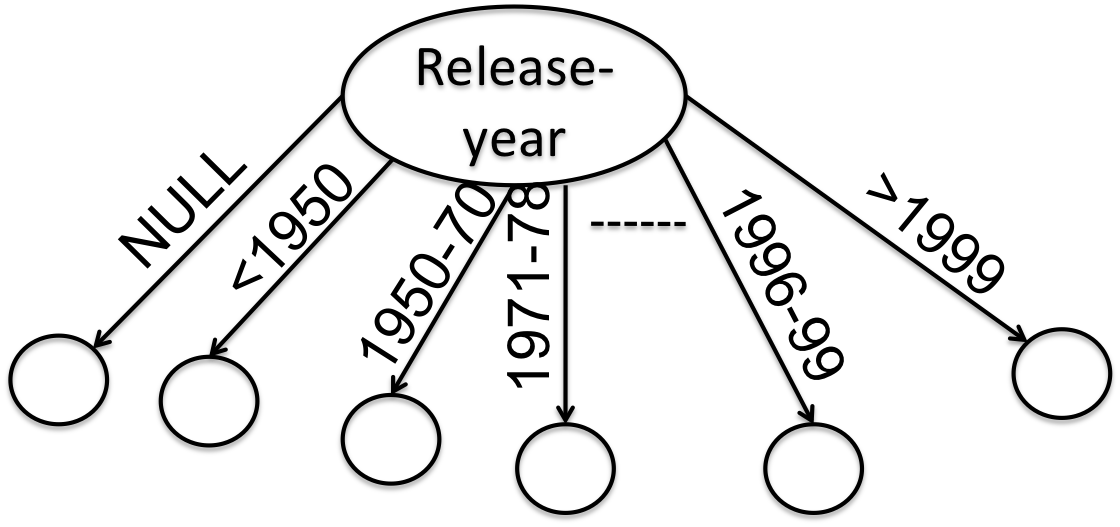}
\vspace{-5mm}
\caption{\label{fig:drilldown}Example of drilling-down on release-year of movies.}
\end{figure}

\vspace{-4mm}
\begin{example}
Figure~\ref{fig:drilldown} gives an example of a hash function that may be obtained using the drill down problem. This hash function may be added to the existing space of hash functions in consideration by a blocking function construction algorithm such as Algorithm~\ref{algo:btree}.
\end{example}

Next we provide an optimal polynomial-time algorithm for the drill down problem based on dynamic programming. We use two core ideas in the algorithm described next. First, suppose we are finding the first partition in the given domain, the only ``interesting endpoints'' of a partition must be either a value at which a duplicate entity lies, or must be due to the boundary caused by the cost bound. Intuitively, we discretize the domain, and now only need to look at a finite number of endpoints in constructing the optimal partition; the space of possible DCC partitions still remains exponential. This observation is formalized below.

\begin{lemma}[Interesting Endpoints]\label{lem:interestingpoints}
Given a domain $D$, $\prec$, with least and greatest elements $start$, $end$, with duplicate pairs ${\cal T}^+ = \{(a_1^1,a_1^2), \ldots, (a_n^1,a_n^2)\}$, cost function $I$ and a cost bound $S$, consider finding the first partition $[start,X)$ or $[start,X]$ (or open interval on $start$ if $start\not\in D$) for the drill down problem. Let $Y\preceq end$ be the greatest value such that $cost([start,Y])\leq S$, then there is an optimal drill down solution with $X\in (\{Y\} \cup \{a_i^j | a_i^j\preceq Y\})$.
\end{lemma}
\eat{
\begin{proof}
(Sketch.) Clearly we cannot have $X\succ Y$. Let us use $IE$ to denote $(\{Y\} \cup \{a_i^j | a_i^j\preceq Y\})$.  Suppose the optimal solution chooses some $X=X^*$, where $X^*\not\in IE$ and $X^*\preceq Y$. Let $X^0$  be the smallest element in $IE$ that is greater than $X^*$. Suppose the next partition chosen by the optimal solution with an endpoint at least as big as $X^0$ is $X^{**}$; we can create an equivalent drill-down solution by creating interval $[start,X^0)$ followed by interval $[X^0,X^{**})$ (or $[X^0,X^{**}]$ as chosen by the optimal solution). Note that our construction is identical to the optimal solution but satisfies $X\in IE$.
\end{proof}
}

\noindent The second observation is the optimal substructure property exploited by our dynamic programming algorithm. Given a domain $D,\prec,start,end$ over which we want to solve the drill down problem, the optimal solution for a sub-domain $D_s,\prec,start',end$, with $start'\succ start$, with the same cost function and cost bound is identical irrespective of the partitions chosen for $D-D_s$, i.e., from $start$ to $start'$. This property allows us to memoize the solutions for all sub-domains of known interesting end-points, namely from $a_i^j$ to $end$, for every $a_i^j$. We can then find an optimal solution to the entire domain by recursively considering sub-domains, as formalized below.

\begin{lemma}[Optimal Substructure]\label{lem:optstructure}
Given a domain $D,\prec,start,end$ with duplicate pairs ${\cal T}^+ =
\{(a_1^1,a_1^2), \ldots, (a_n^1,a_n^2)\}$, cost function $I$, cost bound
$S$, let $Y$ be greatest value satisfying cost bound (as defined in
Lemma~\ref{lem:interestingpoints}). Let $V(I)$ be the total number of
violations in the optimal solution for the subset of ${\cal T}^+$ with each
endpoint in $I$. Then, $V(D)$ can be recursively computed as:
$$V([a,end]) = \min_{P\in (\{Y\}\cup \{a_i^j | a\prec a_i^j \preceq Y\})}
(B([a,P]) + V((P,end]))$$\footnote{A similar expression for $B([a,P)) +
V([P,end])$, which is omitted. We have a similar formula for every combination
of open and closed interval, i.e., $[a,end)$, $(a,end)$, $(a,end]$.}
\noindent where $B([a,P])$ is the number of duplicate pairs broken due to the interval $B([a,P])$; i.e., $B([a,P]) = |\{i | a\preceq a_i^1\preceq P\prec a_i^2\}|$.
\end{lemma}
\eat{
\begin{proof}
(Sketch.) The proof follows from the fact that Lemma~\ref{lem:interestingpoints} gives a complete set of points to consider when starting at $a$, and the above formula considers all these points to find the best $V$. The value of $V$ is given by the total number of violations in ${\cal T}^+$ due to the interval from $a$ to $end$, i.e., function $B()$, and the total number of violations on the rest of the domain.
\end{proof}
}

\noindent The above lemma provides a natural dynamic programming algorithm (described in Algorithm~\ref{algo:drilldown}), where we recursively solve the drill down problem for sub-domains, and memoize these solutions for future recursive calls in $M$; initially, no solution is memoized. Algorithm~\ref{algo:drilldown} returns the total number of violated duplicate pairs but also tracks the specific endpoints. It can be seen easily that this algorithm runs in near-linear time and space based on the observation that the total number of different recursive calls is at most ${\cal O}(n)$: ${\cal O}(n)$ corresponding to all possible endpoints of duplicate pairs, and another ${\cal O}(n)$ corresponding to each maximum $Y$-value from Lemma~\ref{lem:optstructure} for each endpoint.

\begin{algorithm}[t]
{\scriptsize
\caption{Sketch of the dynamic programming algorithm with memoization to solve the drill down problem for a given domain $D$ with a set of duplicate pairs ${\cal T}^+$.}
\begin{algorithmic}[1]
\STATE {\bf Input:} $D=[a,b], {\cal T}^+$, cost function: $\mbox{cost}(\cdot)$, cost bound $S$, memoized solutions $M:D_s\rightarrow NULL \cup \mathbb{N}$
\IF{$M(D)\neq NULL$}
\STATE {\bf Return} $M(D)$.
\ENDIF
\IF{${\cal T}^+=\emptyset$}
\STATE $M(D)=0$. {\bf Return} $M(D)$.
\ENDIF
\STATE Compute max $Y$-value from $a$ using $\mbox{cost}(\cdot)$, $S$ (Lemma~\ref{lem:interestingpoints}).
\STATE Let ${\cal Z} = (\{Y\}\cup \{a_i^j\in {\cal T}^+ | a \prec a_i^j \preceq Y\})$
\STATE Minimum value $m=\infty$, endpoint $p_{opt}=NULL$
\FOR{$P\in {\cal Z}$}
\STATE Compute $B([a,P])$ using ${\cal T}^+$ (Lemma~\ref{lem:optstructure})
\STATE $cand = B([a,P]) + V([P,end])$ // $V([P,end])$ computed recursively
\STATE $M([P,end]) = V([P,end])$
\IF{$m>cand$}
\STATE $m=cand$; $p_{opt}=P$
\ENDIF
\ENDFOR
\STATE $M(D)=m$. {\bf Return} $M(D)$
\end{algorithmic}
\label{algo:drilldown}
}
\end{algorithm}

So far we have considered the drill down problem under the disjointness condition (recall Definition~\ref{defn:dcc}). We finally note that the drill down problem is trivial if we were allowed a non-disjoint set of intervals: We simply look at each duplicate pair $(a_i^1,a_i^2)$ individually and create an interval $I_i=[a_i^1,a_i^2]$ if and only if $\mbox{cost}(I_i)\leq S$.

\section{Non-Disjoint Canopies}
\label{sec:nondisjoint}

In this section we consider the construction of a set of canopies that don't need to be disjoint. The first thing to note is that we need to revise our cost model from Section~\ref{subsec:cost}. We note that a cost function that only penalizes the size of the largest canopy doesn't suffice any longer: Given a set $U$ of entities, we can create $\frac{|U|(|U|-1)}{2}$ canopies, with one canopy for each pair of entities in $U$. Note that this set of canopies has a maximum canopy size of $2$, and a perfect recall of $1$. However, constructing a canopy for each pair is clearly prohibitive, as it incurs a large communication cost, i.e., each entity needs to be transferred to machines corresponding to ${\cal O}(|U|)$ canopies. Therefore, we introduce a cost metric that minimizes the combination of communication and computation cost. The cost of a set ${\cal C}=\{C_1, \ldots, C_m\}$ is given by:
$$\mbox{cost}({\cal C}) = \max_{1\leq i\leq m} |C_i|^2 + \sum_{i=1}^m |C_i|$$
\noindent The computation cost, as before, is approximated by the computation for the largest canopy, where a complete pairwise comparison is performed. The communication is given by the total size of all canopies put together, which is roughly the number of entities that need to be transferred to different machines. 

We address the problem of finding non-disjoint canopies as finding sets of canopies ${\cal C}_1, {\cal C}_2, \ldots$, where each ${\cal C}_i$ is a disjoint set of canopies. In a distributed environment,  each ${\cal C}$ can be performed in one map-reduce round. (Alternatively, if non-disjoint canopies are inherently supported, we may simply construct a single set $\bar {\cal C}$ of canopies as $\bar {\cal C}=\bigcup_{i} {\cal C}_i$.) When treating non-disjoint canopies as multiple rounds of disjoint canopies, once we bound the computation cost (i.e., the size of the largest canopy) in each round, our goal reduces to minimizing the number of rounds to obtain maximum recall with respect to a training dataset.

We present a generic algorithm (Algorithm~\ref{algo:nondisjoint}) that extends any algorithm for disjoint canopy formation to an algorithm for the non-disjoint case. We assume a bound on the maximum computation in any round, and use the disjoint algorithm to maximize recall in a round. The duplicate pairs that are covered are then removed from the labeled dataset, and the next round is performed. We may truncate the algorithm when all pairs are covered, or no more pairs can be covered, or a pre-specified maximum number of rounds has reached.

\begin{algorithm}[t]
{\scriptsize
\caption{Generic algorithm for performing non-disjoint canopy formation as multiple rounds of disjoint canopy formation.}
\begin{algorithmic}[1]
\STATE {\bf Input:} Labeled data ${\cal T}^+$, maximum canopy-size bound $S$, disjoint-algorithm {\sc AlgoDisj} returning the covered pairs, (optional) bound on the number of rounds $R$.
\STATE {\em numRds}$=0$, {\em change}$=\mbox{true}$
\WHILE{$({\cal T}^+\neq \emptyset) \wedge (\mbox{change}) \wedge (\mbox{numRds}<R)$}
\STATE {\em numRds}$=${\em numRds}$+1$; {\em change}$=\mbox{false}$
\STATE {\sc Covered}={\sc AlgoDisj}$({\cal T}^+,S)$
\IF{{\sc Covered}$\neq \emptyset$}
\STATE ${\cal T}^+ = {\cal T}^+ \setminus \mbox{\sc Covered}$
\STATE {\em change}$=\mbox{true}$
\ENDIF
\ENDWHILE
\end{algorithmic}
\label{algo:nondisjoint}
}
\end{algorithm}

\newcommand{\algorand}{R}
\newcommand{\algosingle}{SH}
\newcommand{\algochain}{C}
\newcommand{\algochaintree}{CT}
\newcommand{\algotree}{HBT}
\newcommand{\algobilenko}{BKM}

\section{Experiments}
\label{sec:experiments}

This  section presents a detailed experimental study using two large commercial datasets at Yahoo: (1) a movie dataset consisting of 140K entities, and (2) a restaurants dataset consisting of 40K entities. We present a summary of results based on both the datasets, but focus on movies for a more detailed evaluation. (We focus only on one dataset for a detailed evaluation due to space constraints; the movies dataset being larger makes for a more interesting study although trends are similar in the restaurants dataset.)

The primary goal of our study is to measure the effectiveness (increased recall) due to the more expressive \btree-based blocking, as compared to restrictions of \btree s. We measure recall for disjoint and non-disjoint versions of all our algorithms. In addition to the primary objectives described above, our experiments also understand the effects of increasing the size of canopies on recall, variation of recall with the number of disjuncts, effects of specific greedy strategies used, and understanding some basic properties of \btree-based blocking. Our experimental setup is described in Section~\ref{subsec:setup} and results are presented in Section~\ref{subsec:expresults}.

\eat{
\begin{enumerate}
\item For disjoint blocking (i.e., one map-reduce round), we evaluate the effectiveness of our \btree-based blocking approach, as compared to previously proposed techniques, and other restrictions of \btree; 
\item Under non-disjoint blocking, we measure recall for each blocking method as the number of rounds is increased;
\item Measure the effectiveness of rollup applied to small canopies (generated using any of the approaches).
\end{enumerate}
}

\subsection{Experimental Setup}
\label{subsec:setup}

\begin{table*}[t]
\begin{center}
{\scriptsize
\begin{tabular}{|c|c|}
\hline
{\bf Attributes} & {\bf Hash function} \\ \hline
All & (1) $h(x)=x$; (2) $h(x)=$ prefix/suffix of length $K$; (3) $h(x)=$ most frequent $K$ characters in alphanumeric order;  ($K=1,3,5$)\\ \hline
{\tt title} & $h(x)=$longest token of $x$\\ \hline
{\tt year}, {\tt runtime} & number rounded to nearest to create $k$-point intervals, i.e., $h(x)=x-(x \mod k)$\\ \hline
{\tt director} & (1) $h(x)=$first-name of $x$; (2) $h(x)=$last-name of $x$\\ \hline
\end{tabular}
}
\caption{\label{tab:hashspace} Sample of the space of hash functions on movies used in our experiments. }
\end{center}
\end{table*}

\subsection*{Dataset}

We have applied \cobbler\ on two commercial datasets from a search engine company: {\tt movies} and {\tt restaurants}. The primary movies dataset used in our experiments is a large database $D_{movie}$ of 140K movies from Yahoo. In addition, we use a sample of movies from DBPedia~\cite{dbpedia} to obtain new duplicates, in addition to the duplicate already existing in $D_{movie}$. We constructed a labeled dataset ${\cal T}^+$ consisting of $1054$ pairs of duplicates: Around $350$ pairs of duplicates were obtained using manual labeling by paid editors. The remaining $704$ pairs were obtained automatically by finding common references to IMDb~\cite{IMDb} movies; a small sample of $100$ automatically generated pairs were checked manually to confirm  that these were all duplicates. \eat{Finally, we also generated a large set ${\cal T}^-$ of $1500$ negative pairs, once again using a combination of manual and automatic approaches. (Note that ${\cal T}^-$ is not used by any of our techniques, but we used them in evaluating the algorithm from~\cite{bilenko06:blocking}.)} The schema of movies consisted of attributes {\tt title}, {\tt director}, {\tt release year}, {\tt runtime}, {\tt genre} on which hash functions were created, and also other attributes (such as {\tt genre} and {\tt crew members}) that weren't used for blocking. A sample of the space of hash functions used in our experiments is shown in Table~\ref{tab:hashspace}.

The restaurants dataset used in our experiments consists of $40,000$ restaurant records with attributes {\tt name}, {\tt street}, {\tt city}, {\tt state}, and {\tt zip}. After de-duplication, there are $13,000$ unique restaurant records. We use a labeled dataset of $4,674$ duplicate pairs, and we used a similar set of hash functions as in Table~\ref{tab:hashspace}.

\subsection*{Metrics}
We evaluate our canopy generation algorithms using two metrics -- {\em  recall} and {\em computation cost}. Recall is measured as the fraction of matching pairs in ${\cal T}^+$ that appear within some canopy (Definition~\ref{defn:labeled}).
Our algorithms are used to learn blocking hash functions, which are in turn applied to new data. We measure the computation cost in terms of the time taken to apply the hash function learnt by our algorithms on the dataset. Note that this is {\em not} the time taken to learn blocking functions. For non-disjoint canopy formation (Algorithm~\ref{algo:nondisjoint} in Section~\ref{sec:nondisjoint}) we measure the increase in recall as the number of disjuncts (or map-reduce steps) is increased.

\subsection*{Algorithms}
We describe our algorithms next. If any of our algorithms result in canopies $C$ with size larger than our maximum size limit $S$, we further split it {\em randomly} into $\lceil \frac{|C|}{S}\rceil$ smaller parts. The algorithms we compare are
\squishlist
\item Random (\algorand): Each entity in ${\cal U}$ is assigned uniformly at random to one of $\lceil \frac{|{\cal U}|}{S}\rceil$ canopies.
\item Single-Hash (\algosingle): Canopies are formed by picking a single hash function which maximizes recall.
\item Chain (\algochain): Canopies are formed by picking the best conjunction of hash function. (Note that this is the ``size-aware'' analogue of the approaches taken by previous work~\cite{bilenko06:blocking,knoblock} on using labeled data.)
\item Chain-Tree (\algochaintree): A restriction to our tree based hash function where the same hash is used at each level. \algosingle, \algochain, \algochaintree\ were described earlier in Section~\ref{sec:restricted}.
\item Hierarchical Blocking Tree (\algotree): Our \btree-based canopy generation algorithm presented in Algorithm~\ref{algo:btree}.
\squishend

We also consider non-disjoint variants of all our algorithms; if $A \in \{\mbox{\algosingle, \algochain, \algochaintree, \algotree}\}$ denotes one of the above algorithms, we use $A\mbox{-ND}$ to denote its non-disjoint variant (i.e., using $A$ in Algorithm~\ref{algo:nondisjoint}). 

\subsection*{Setup}

We perform 5-fold cross-validation for all runs of algorithms: We split ${\cal T}^+$ into 5 equal pieces randomly, then average over five runs with each run using $4$ pieces of ${\cal T}^+$ as a training set to obtain the blocking function, then use the $5$th piece as a test set. Since we don't make any novel contribution on size estimation, our oracle $size()$ computes the exact sizes of canopies based on the entire dataset. Our experiments were performed by varying the allowed maximum canopy size with $1K, 5K, 10K, 20K, 100K$ entities per canopy.

\subsection{Results}
\label{subsec:expresults}

We start by presenting detailed results on the movie dataset (Section~\ref{subsubsec:m1}--\ref{subsubsec:m2}). Finally, we present a brief summary of results on the restaurants dataset in Section~\ref{subsubsec:rest}.

\subsubsection{Disjoint Canopies}
\label{subsubsec:m1}

\begin{figure*}[t]
\centering
\subfigure[Comparison of disjoint canopy formation algorithms, varying the maximum size of canopies.]{\includegraphics[width=2in]{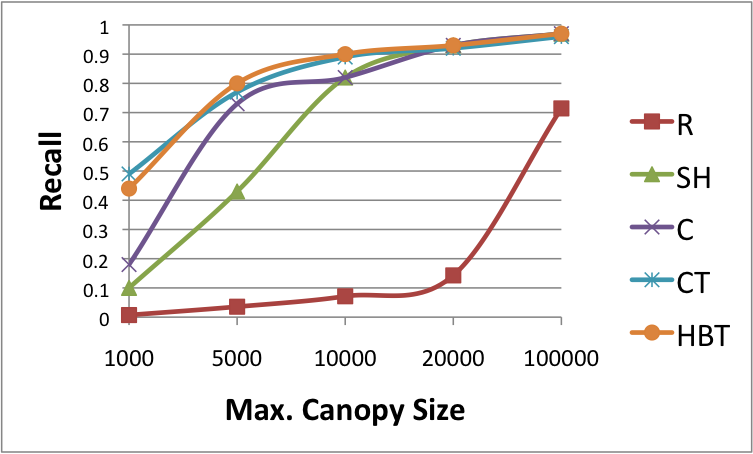}\label{fig:overall-recall}}
\subfigure[Comparison of optimistic, pessimistic, and expected greedy picking strategies for each algorithm, fixing the maximum canopy size to 10000.]{\includegraphics[width=2in]{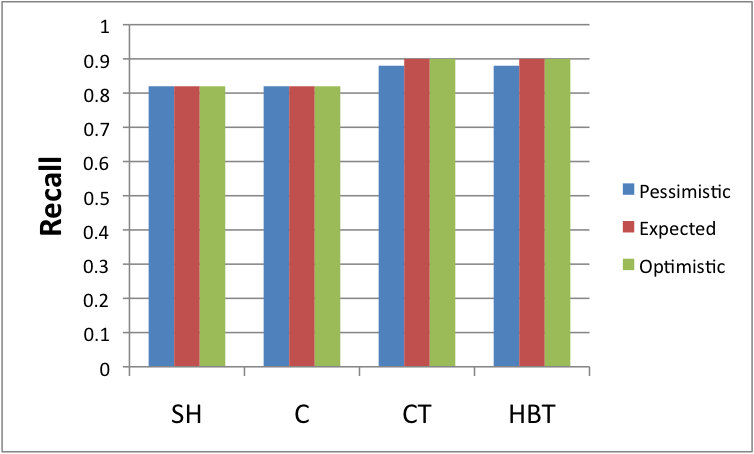}\label{fig:opt-pess-exp}}
\subfigure[Comparison of overall recall for non-disjoint canopy formation algorithms, varying the maximum size of canopies.]{\includegraphics[width=2in]{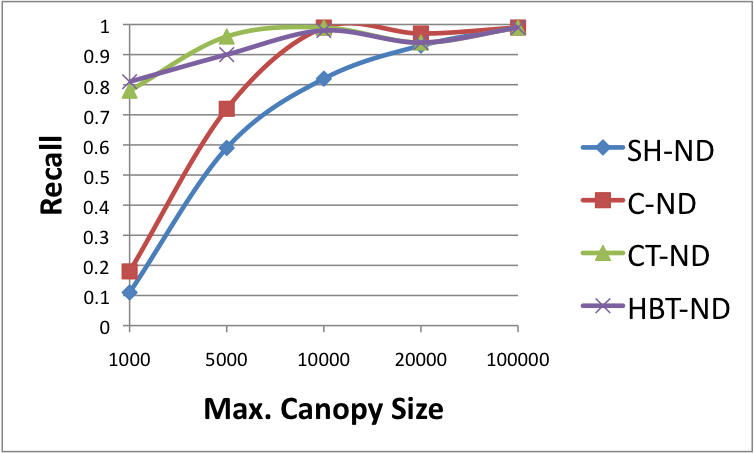}\label{fig:nondisjoint-overall}}
\subfigure[Benefit of performing non-disjoint by comparing \algotree-ND and \algotree\ varying maximum canopy size.]{\includegraphics[width=1.8in]{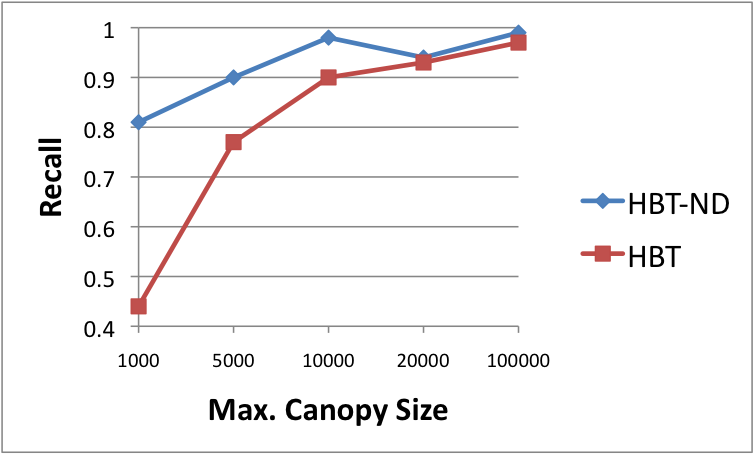}\label{fig:disjoint-nondisjoint}}
\subfigure[Variation of recall as the number of rounds is increased, for all non-disjoint canopy formation algorithms with maximum canopy size $5000$.]{\includegraphics[width=1.8in]{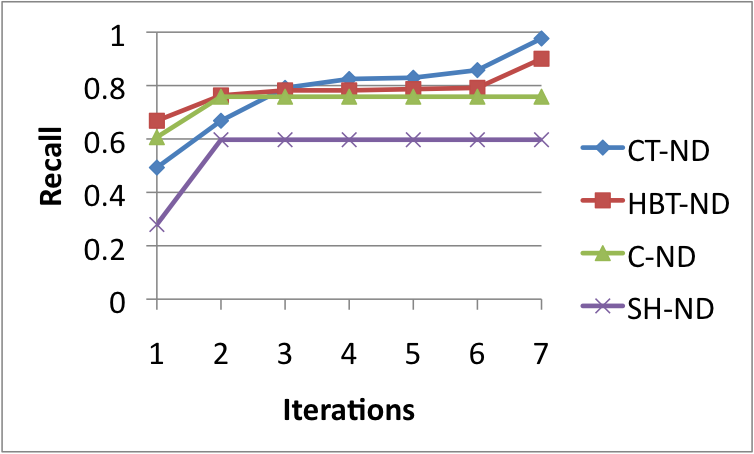}\label{fig:iterations}}
\subfigure[Summary of recall on applying \cobbler\ on the restaurants dataset, varying the maximum allowed canopy size. Comparison of \btree-based and conjunctive blocking.]{\scriptsize
\begin{tabular}{|c|c|c|c|c|c|}
\hline
{\bf Approach} & {\bf Algorithm} & {\bf 100} &  {\bf 200} &  {\bf 500} &  {\bf 1000} \\ \hline
\multirow{2}{*}{{\bf Disjoint}}  & {\bf \algochain} & $0.33$ & $0.49$ & $0.66$ & $0.75$ \\ \cline{2-6}
& {\bf \algotree} &  $0.84$ & $0.87$ & $0.89$ & $0.91$\\ \hline
\multirow{2}{*}{{\bf Non-disjoint}}  & {\bf \algochain-ND} & $0.38$ & $0.51$ & $0.66$ & $0.75$ \\ \cline{2-6}
& {\bf \algotree-ND} &  $0.97$ & $0.98$ &  $0.99$ & $0.99$ \\ \hline
\end{tabular}
\label{tab:restaurants}}
\caption{Experimental Results}
\end{figure*}

Our first experiment was to compare the overall recall obtained by each of the algorithms---\algorand, \algosingle, \algochain, \algochaintree, and \algotree. Figure~\ref{fig:overall-recall} shows the recall obtained by each of the algorithms on the movie dataset, varying the maximum allowed canopy size. (For each of the algorithms, we picked the best of the optimistic, pessimistic, and expected greedy picking strategies.) The most important observation is that \algotree\ achieves a significantly higher recall than \algochain\ and \algosingle, particularly when the maximum canopy size is lower. The reason for \algotree's higher recall is the greater expressive power of \btree s as a construct for describing disjoint blocking functions; \btree's are able to apply a hash function at the first level that creates many good small canopies and a few large canopies, which are further split at subsequent levels of the tree.  Another interesting observation from Figure~\ref{fig:overall-recall} is that \algochaintree\ performs roughly as well as \algotree, despite the slightly lower expressive power: Intuitively, the added power of \algotree\ is effective when different nodes in the same level need different hash functions. Such a case would arise when different sections of the data have differing properties (e.g., US movies versus German movies); our dataset, however, only contained US movies.
Finally, as expected \algorand\ gives the lowest recall among all algorithms; henceforth, we omit \algorand\ from the rest of our experiments.

\eat{
\begin{table}[t]
\begin{center}
{\small
\begin{tabular}{|c|c|c|c|}
\hline
{\bf Function} & {\bf \algosingle} & {\bf \algochaintree}  & {\bf \algotree} \\ \hline
{\bf Time per record (ms)} & $8.8$  & $17.7$ &  $7.1$ \\ \hline
{\bf Average tree height} & 1 & $2.8$ &  $1.34$ \\ \hline
\end{tabular}
}
\caption{\label{tab:timing} (1) Average running time (in $\mu s$.) of applying the best blocking function for each record. (2) Average length of the tree. All the numbers are for a max canopy size of $10,000$.}
\vspace{-0.5cm}
\end{center}
\end{table}
}

To further understand the effects of the three greedy picking strategies---optimistic, pessimistic, and expected---described in Section~\ref{subsec:btreegreedy}, in Figure~\ref{fig:opt-pess-exp} we plot the recall for each of the algorithms by varying the greedy picking strategy. We note that in most cases all three algorithms perform very similarly, with the optimistic picking strategy slightly outperforming the others. The intuition for optimistic greedy strategy performing slightly better is that an optimistic estimate is better than an expected estimate since future levels of blocking are significantly better than a random split of each large canopy. Since optimistic is never worse than the other strategies, for the rest of our experiments we choose the optimistic strategy for each algorithm.

\eat{
\begin{figure}[t]
\centering
\includegraphics[width=2.8in]{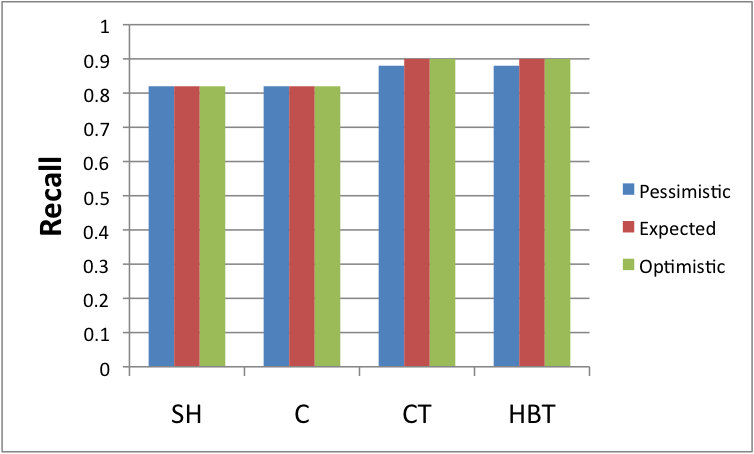}
\caption{\label{fig:opt-pess-exp}Comparison of optimistic, pessimistic, and expected greedy picking strategies for each algorithm, fixing the maximum canopy size to 10000.}
\end{figure}
}

\subsubsection{Non-disjoint Canopies}

Next we consider the non-disjoint variants of each of the algorithms.  Figure~\ref{fig:nondisjoint-overall} shows the overall recall for each non-disjoint algorithm as the number of canopies is varied. We notice, once again, that \algotree-ND achieves a significantly higher recall than \algochain-ND and \algosingle-ND. In particular, \algochain-ND is the size-aware analogues of previous state of the art~\cite{bilenko06:blocking,knoblock}. The reason for a much higher recall in \algotree-ND is again the larger space of blocking functions \btree s can represent. Specifically, any conjunction that contains even one canopy larger than the maximum allowed is not permitted (or the conjunction needs to be further restricted losing more duplicate pairs).  Note however that overall recall of \algochaintree-ND is very similar to that of \algotree-ND; however, we shall see shortly that in the initial rounds of disjunction, \algotree-ND increases recall slightly more rapidly than \algochaintree-ND.

\eat{
\begin{figure}[t]
\centering
\includegraphics[width=2.8in]{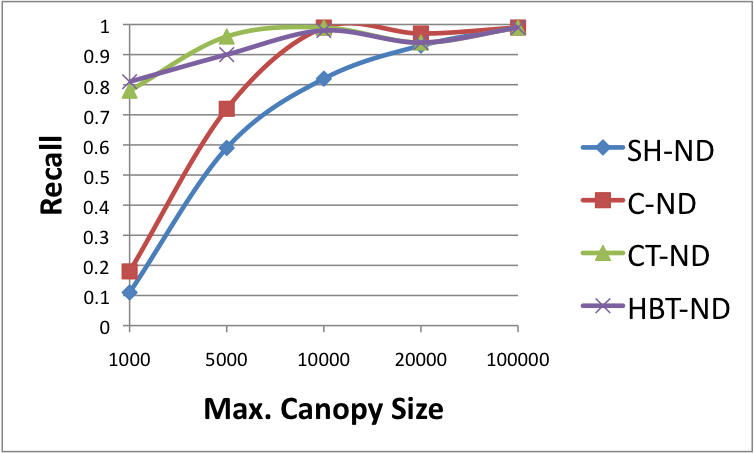}
\caption{\label{fig:nondisjoint-overall}Comparison of overall recall for non-disjoint canopy formation algorithms, varying the maximum size of canopies.}
\end{figure}
}

A second observation on non-disjoint canopy formation is that the non-disjoint versions of each algorithm obtain higher recall than the corresponding disjoint versions. In Figure~\ref{fig:disjoint-nondisjoint} we show the increase in recall obtained by \algotree-ND as compared to \algotree, for each maximum canopy size. Note that the additional benefit of non-disjointness diminishes as the maximum canopy size is increased.

\eat{
\begin{figure}[t]
\centering
\includegraphics[width=2.8in]{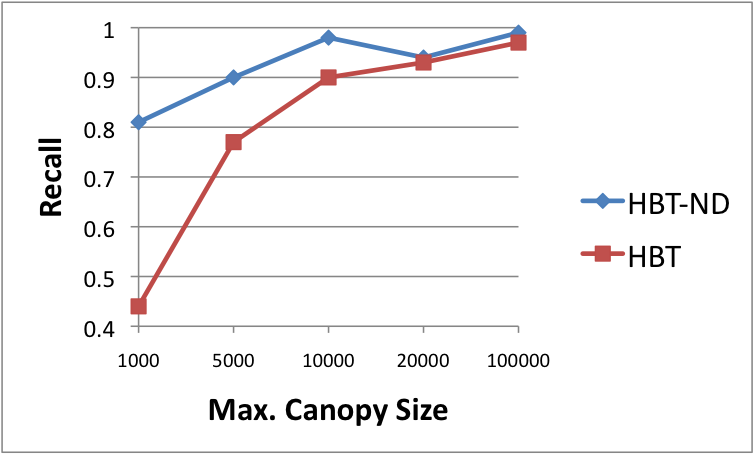}
\caption{\label{fig:disjoint-nondisjoint}Benefit of performing non-disjoint by comparing \algotree-ND and \algotree\ varying maximum canopy size.}
\end{figure}
}

Next let us take a closer look at how the recall changes as the number of iterations is increased. To examine the difference between \algochaintree-ND and \algotree-ND (as well as other non-disjoint algorithms), we plot the recall obtained after each round of disjoint canopy formation. Figure~\ref{fig:iterations} plots the overall recall for the case of maximum canopy size $5000$; we picked one fold of our cross-validation in which \algochaintree-ND ends with a slightly higher recall than \algotree\ (therefore the apparent discrepancy with Figure~\ref{fig:overall-recall}). First, note that for every iteration, \algotree-ND is better than \algochain-ND and \algosingle-ND, which means that the number of positive examples covered increases more steeply for \algotree-ND. Second, we see that \algotree-ND obtains a higher recall than \algochaintree-ND initially, but \algochaintree-ND eventually ends with a slightly higher recall; in other words, with a limited number of map-reduce rounds, \algotree-ND performs better than \algochaintree-ND. An optimal strategy of choosing a non-disjoint canopy formation by combining \algotree-ND and \algochaintree-ND is left as future work.

\eat{
\begin{figure}[t]
\centering
\includegraphics[width=2.8in]{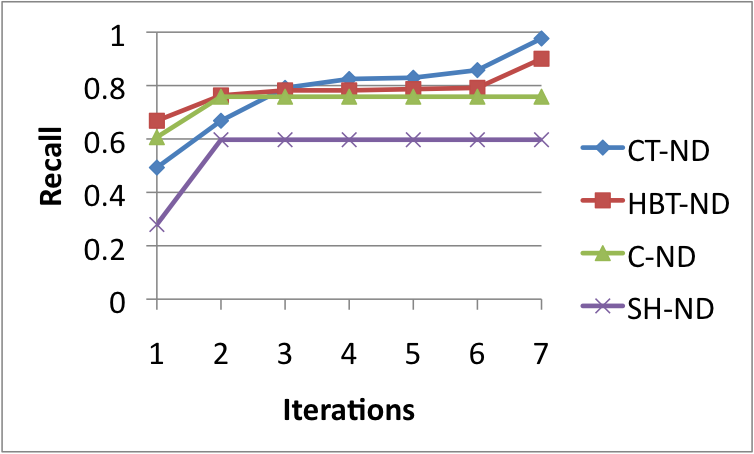}
\caption{\label{fig:iterations}Variation of recall as the number of rounds is increased, for all non-disjoint canopy formation algorithms with maximum canopy size $5000$.}
\end{figure}
}



\subsubsection{Computation cost and Tree size}
\label{subsubsec:m2}

\ \newline \noindent {\bf Computational Cost:} We compared the computation cost (i.e., running) of applying a \btree\ against the cost of applying other blocking functions. The primary objective of investigating the running time is to establish the fact that \btree s do not add significant burden on the time required to apply the blocking function on an entire dataset. Table~\ref{tab:timing} shows the running time of applying the best blocking function (for maximum canopy size $10,000$) for each of the algorithms (conjunctions being similar to applying a single hash function are omitted); these numbers are averaged over the $\sim 140K$ movie entities and over 5 repeated applications of the blocking function on the entire dataset. We note that applying each of the blocking functions requires a negligible amount of time (always under $20 \mu s$ per record), and \btree s don't add any discernible computational cost. 

\ \newline \noindent {\bf Tree Size:} Table~\ref{tab:timing} also shows the height of the tree for \algochaintree\ and \algotree\ (averaged over the $5$ folds of cross-validation). It is noteworthy that \algotree\ obtains similar recall with a shorter \btree\ than \algochaintree. This is because the \btree\ constructed using \algotree\ is able to selectively create longer branches only when necessary. The longer tree for \algochaintree\ explains the higher blocking time per record.

\begin{table}[t]
\begin{center}
{\small
\begin{tabular}{|c|c|c|c|}
\hline
{\bf Function} & {\bf \algosingle} & {\bf \algochaintree}  & {\bf \algotree} \\ \hline
{\bf Time per record (ms)} & $8.8$  & $17.7$ &  $7.1$ \\ \hline
{\bf Average tree height} & 1 & $2.8$ &  $1.34$ \\ \hline
\end{tabular}
}
\caption{\label{tab:timing} (1) Average running time (in $\mu s$.) of applying the best blocking function for each record. (2) Average length of the tree. All the numbers are for a max canopy size of $10,000$.}
\vspace{-0.5cm}
\end{center}
\end{table}

\subsubsection{Summary of Results for Restaurants}
\label{subsubsec:rest}

We present a very brief summary of our results on the restaurants dataset; restaurants displayed a similar general trend as movies, and a detailed study of restaurants is omitted due to space constraints. Table~\ref{tab:restaurants} presents the overall recall for \algotree\ and \algotree-ND compared against \algochain\ and \algochain-ND, varying the sizes of the maximum canopy: (1) We note that both the disjoint and non-disjoint versions of \algotree\ significantly outperform the disjoint and non-disjoint versions of conjunctive blocking. (2) Further, as with movies, the recall achieved by \algotree\ is very high on restaurants, and very close to $1$ with non-disjoint blocking even for small canopy sizes.

{\scriptsize
\bibliographystyle{abbrv}
\bibliography{canopy}
}




\end{document}